\documentclass[pra,aps,twocolumn,superscriptaddress,showpacs]{revtex4}
\usepackage[dvips]{graphicx}
\begin{document}
\author{Jing-Ling Chen}
\email{chenjl@nankai.edu.cn} \affiliation{ Liuhui Center for Applied
Mathematics and Theoretical Physics Division, Chern Institute of
Mathematics, Nankai University, Tianjin 300071, People's Republic of
China}
\author{Kang Xue}
\affiliation{Department of Physics, Northeast Normal University,
Changchun, Jilin 130024, People's Republic of China}
\author{Mo-Lin Ge}
\email{geml@nankai.edu.cn} \affiliation{ Liuhui Center for Applied
Mathematics and Theoretical Physics Division, Chern Institute of
Mathematics, Nankai University, Tianjin 300071, People's Republic of
China}

\title {Braiding transformation, entanglement swapping and Berry phase in entanglement space}
\begin{abstract}
We show that braiding transformation is a natural approach to
describe quantum entanglement, by using the unitary braiding
operators to realize entanglement swapping and generate the GHZ
states as well as the linear cluster states. A Hamiltonian is
constructed from the unitary
$\check{R}_{i,i+1}(\theta,\varphi)$-matrix, where $\varphi=\omega t$
is time-dependent while $\theta$ is time-independent. This in turn
allows us to investigate the Berry phase in the entanglement space.
\end{abstract}

 \maketitle

\newcommand{\half}{{\textstyle \frac{1}{2}}}
\newcommand{\mn}{\mu\nu}
\newcommand{\eps}{\varepsilon}
\renewcommand{\L}{\mathcal{L}}

\newcommand{\eq}{\begin{equation}}
\newcommand{\en}{\end{equation}}
\newcommand{\eqa}{\begin{eqnarray}}
\newcommand{\ena}{\end{eqnarray}}

\def\l{\langle}
\def\r{\rangle}



\section{Introduction}
Quantum entanglement is the most surprising nonclassical property of
composite quantum systems that Schr{\" o}dinger singled out many
decades ago as ``the characteristic trait of quantum mechanics".
Recently entanglement has become one of the most fascinating topics
in quantum information, because it has been shown that entangled
pairs are more powerful resources than the separable ones in a
number of applications, such as quantum cryptography \cite{Ekert},
dense coding, teleportation \cite{Benne1} and investigation of
quantum channels, communication protocols and computation
\cite{Niels}. For instance, by using a maximally entangled state
$|\Phi^+\rangle = 1/\sqrt{2} (|\uparrow \uparrow \rangle
+|\downarrow \downarrow\rangle)$ (i.e., one of Bell states and also
the so-called Einstein-Podolsky-Rosen (EPR) channel in
\cite{Benne1}), Bennett \emph{et al.} have showed that it is
faithful to transmit a one-qubit state $a |\uparrow\rangle+ b
|\downarrow\rangle$ from one location (Alice) to another (Bob) by
sending two bits of classical information.

For a two-qubit system, there has been defined a ``magic basis"
consisting of
four Bell states \cite{magic}:
\begin{eqnarray}\label{magic}
&& |\Phi^+\rangle = 1/\sqrt{2}
(|\uparrow \uparrow \rangle +|\downarrow \downarrow\rangle),\nonumber\\
&& |\Phi^-\rangle = 1/\sqrt{2}
(|\uparrow \uparrow \rangle -|\downarrow \downarrow\rangle),\nonumber\\
&& |\Psi^+\rangle = 1/\sqrt{2}
(|\uparrow \downarrow \rangle +|\downarrow \uparrow\rangle),\nonumber\\
&& |\Psi^-\rangle = 1/\sqrt{2} (|\uparrow \downarrow \rangle
-|\downarrow \uparrow\rangle),
\end{eqnarray}
where spin-1/2 notation for definiteness has been used. Any pure
state of two-qubit can be expanded in this particular basis and its
degree of entanglement can be expressed in a remarkably simple way
\cite{magic}. It is possible to study these Bell states from the
other point of view of transformation theory. The fact that they are
all normalized and mutual orthogonal naturally indicates that the
four Bell states are connected to the standard basis
$\{|\uparrow\uparrow\rangle,|\uparrow\downarrow\rangle,|\downarrow\uparrow\rangle,
|\downarrow\downarrow\rangle\}$ by a unitary transformation
\eq \label{unitary} U = \frac{1}{\sqrt{2}}\left( \begin{array}{cccc} 1 & 0 & 0 & 1 \\ 0 & 1 & 1 & 0 \\
0 & -1 & 1 & 0 \\ -1 & 0 & 0 & 1 \end{array} \right).\en More
precisely, let $|\uparrow\rangle=(1, 0)^T$ and
$|\downarrow\rangle=(0, 1)^T$, $|\uparrow\uparrow\rangle$ is
understood as $|\uparrow\rangle \otimes |\uparrow\rangle$, one then
has the matrix forms for the standard basis as
$|\uparrow\uparrow\rangle=(1,0,0,0)^T$,
$|\uparrow\downarrow\rangle=(0,1,0,0)^T$,
$|\downarrow\uparrow\rangle=(0,0,1,0)^T$,
$|\downarrow\downarrow\rangle=(0,0,0,1)^T$. Acting the unitary
matrix $U$ on the standard basis will produce the four Bell states:
$U |\uparrow\uparrow\rangle=1/\sqrt{2} (1,0,0,-1)^T=|\Phi^-\rangle$,
$U |\uparrow\downarrow\rangle=1/\sqrt{2}
(0,1,-1,0)^T=\bar{|\Psi^-\rangle}$, $U
|\downarrow\uparrow\rangle=1/\sqrt{2} (0,1,1,0)^T=|\Psi^+\rangle$,
$U |\downarrow\downarrow\rangle=1/\sqrt{2}
(1,0,0,1)^T=|\Phi^+\rangle$, in short one obtains $U
(|\uparrow\uparrow\rangle,|\uparrow\downarrow\rangle,|\downarrow\uparrow\rangle,
|\downarrow\downarrow\rangle) = (|\Phi^-\rangle, |\Psi^-\rangle,
|\Psi^+\rangle, |\Phi^+\rangle)$.

During the investigation of the relationships among quantum
entanglement, topological entanglement and quantum computation,
Kauffman \emph{et al.} have discovered a very significant result
that the matrix $U$ is nothing but a braiding operator, and
furthermore it can be identified to the universal quantum gate
(i.e., the CNOT gate) \cite{Kauffman}\cite{Kauffman1}. There is an
earlier literature on topological quantum computation and which is
all about quantum computing using braiding \cite{Kitaev}. These
literatures introduce the braiding operators and Yang--Baxter
equations to the field of quantum information and quantum
computation, and also provide a novel way to study the quantum
entanglement.

Our aim in this work is twofold: one is to show that braiding
transformation is a natural approach describing the quantum
entanglement, the other is to investigate the Berry phase in the
entanglement space (or the Bloch space). The paper is organized as
follows. In Sec. II, we briefly review the unitary braiding
operators and apply them to realize entanglement swapping and to
generate the Greenberger-Horne-Zeilinger (GHZ) states as well as the
linear cluster states. In Sec. III, after briefly reviewing the
Yang--Baxterization approach, we construct a Hamiltonian from the
unitary $\check{R}_{i,i+1}(\theta,\varphi)$-matrix, where $\varphi$
is time-dependent while $\theta$ is time-independent. This in turn
allows us to investigate the Berry phase in the entanglement space.
Conclusion and discussion are made in the last section.

\section{Braiding transformation and its applications}

Hereafter for convenience, we shall denote the spin up
$|\uparrow\rangle$ and down $|\downarrow\rangle$ as $|0\rangle$ and
$|1\rangle$, respectively. Braiding operators are the
generalizations of the usual permutation operators. For $N$ spin-1/2
particles, the permutation operator for the particles $i$ and $i+1$
reads
 \eq \label{permutation} P_{i,i+1} =
\frac{1}{2}(1+{\vec \sigma}_i\cdot {\vec \sigma}_{i+1}) =
 \left( \begin{array}{cccc} 1 & 0 & 0 & 0 \\ 0 & 0 & 1 & 0 \\
0 & 1 & 0 & 0 \\ 0 & 0 & 0 & 1 \end{array} \right),\en Here
$P_{i,i+1}$ is understood as ${\bf 1}_1\otimes{\bf
1}_2\otimes\cdots\otimes{\bf 1}_{i-1}\otimes (1+{\vec \sigma}_i\cdot
{\vec \sigma}_{i+1})/2\otimes{\bf 1}_{i+2}\otimes \cdots\otimes{\bf
1}_{N}$, where ${\bf 1}$ is the $2 \times 2$ unit matrix. The
permutation operator $P_{i,i+1}$ exchanges the spin state
$|k\rangle_i\otimes|l\rangle_{i+1}$ to be
$|l\rangle_i\otimes|k\rangle_{i+1}$.

The braiding operators satisfy the following braid relations: \eqa
\label{braid}
  b_{i, i+1}  b_{i+1, i+2} b_{i, i+1} & = & b_{i+1, i+2} b_{i, i+1} b_{i+1, i+2},
\; i \leq N-2, \nonumber\\
  b_{i, i+1} b_{j, j+1} & = & b_{j, j+1} b_{i, i+1}, \;\;
   | i - j | \ge 2. \ena
The usual permutation operator $P_{i,i+1}$ is a solution of Eq.
(\ref{braid}) with the constraint $P_{i,i+1}^2 = 1$. Physics prefers
to the unitary transformations. One may observe that both $U$ and
$P_{i,i+1}$ are unitary. Two more general unitary braiding
transformations satisfying the braiding relations are
\eq \label{bmBP}B_{i,i+1} = \frac{1}{\sqrt{2}}\left( \begin{array}{cccc} 1 & 0 & 0 & e^{-i\varphi} \\ 0 & 1 & 1 & 0 \\
0 & -1 & 1 & 0 \\ -e^{i\varphi} & 0 & 0 & 1 \end{array}
\right),\nonumber\\\en
\eq {\cal P}_{i,i+1} = \left( \begin{array}{cccc} e^{i\xi_{00}} & 0 & 0 & 0 \\ 0 & 0 & e^{i\xi_{10}} & 0 \\
0 & e^{i\xi_{01}} & 0 & 0 \\ 0 & 0 & 0 & e^{i\xi_{11}} \end{array}
\right),\en which allow additional phase factors. Braiding operators
$B_{i,i+1}$ and ${\cal P}_{i,i+1}$ transform the direct-product
states $|kl\rangle\equiv|k\rangle_i\otimes|l\rangle_{i+1}$ in the
following way
 \begin{equation} \label{bmB}
B_{i,i+1} \left(\begin{array}{c} |0 0\rangle \\| 0 1 \rangle \\
|1 0\rangle \\ |1 1\rangle
\end{array}\right)=\frac 1 {\sqrt{2}}
 \left(\begin{array}{c} |0 0\rangle-e^{i\varphi}|1 1\rangle \\
 |0 1 \rangle - |1 0 \rangle \\ |0 1\rangle + | 1 0 \rangle\\
 e^{-i\varphi} |0 0\rangle+|11\rangle
\end{array}\right)
,\nonumber\\
 \end{equation}
 \eq
{\cal P}_{i,i+1} \left(\begin{array}{c} |0 0\rangle \\| 0 1 \rangle \\
|1 0\rangle \\ |1 1\rangle
\end{array}\right)=
 \left(\begin{array}{c} e^{i\xi_{00}}|0 0\rangle \\
 e^{i\xi_{10}}|10\rangle \\ e^{i\xi_{01}}|01\rangle\\
 e^{i\xi_{11}}|1 1\rangle
\end{array}\right).
\en They may generate entangled states from disentangled ones: (i)
The braiding matrix $B_{i,i+1}$ yields directly the four Bell states
$|\Phi^\pm\rangle$ and $|\Psi^\pm\rangle$ with the relative phase
factor $e^{-i\varphi}$. The phase factor $e^{-i\varphi}$ originates
from the $q$-deformation of the braiding operator $U$ with
$q=e^{-i\varphi}$ \cite{Jimbo}\cite{Slingerland}, and $\varphi$ may
have a physical significance of magnetic flux \cite{Zeilinger}. In
the next section, we shall vary adiabatically the parameter
$\varphi$ to obtain the Berry phase in the entanglement space. (ii)
When ${\cal P}_{i,i+1}$ acts on an initial separable state
$1/\sqrt{2}(|0\rangle +|1\rangle)_i\otimes 1/\sqrt{2}(|0\rangle
+|1\rangle)_{i+1}$, it produces an entangled state $(e^{i\xi_{00}}|0
0\rangle+e^{i\xi_{01}}|0 1\rangle+e^{i\xi_{10}}|1
0\rangle+e^{i\xi_{11}}|11\rangle)/2$ whose degree of entanglement
equals to $|e^{i(\xi_{00}+\xi_{11})}-e^{i(\xi_{01}+\xi_{10})}|/2$.
Thus it is indeed a very natural way for the braiding operators to
describe and to generate quantum entanglement. To strengthen such a
viewpoint, we would like to provide two explicit examples as
applications of braiding transformations as follows.

\begin{figure}
  \includegraphics[width=4cm]{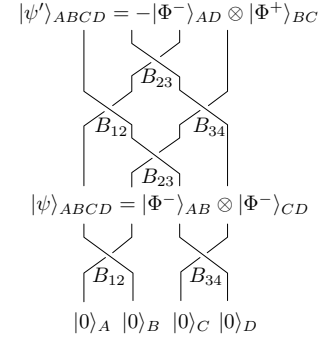}\\
\caption{Realizing ES by braiding transformations. After acting
$B_{34}B_{12}$ on a separable state $|0000\rangle_{ABCD}$, one
prepares a state $|\psi\rangle_{ABCD} =|\Phi^-\rangle_{AB} \otimes
|\Phi^-\rangle_{CD}$ needed for quantum entanglement swapping. After
performing successive braiding transformations
$B_{23}B_{34}B_{12}B_{23}$ on $|\psi\rangle_{ABCD}$, the
entanglement involved in the state $|\psi\rangle_{ABCD}$ is swapped
to the state $|\psi'\rangle_{ABCD}=-|\Phi^-\rangle_{AD} \otimes
|\Phi^+\rangle_{BC}$. }  \label{swapfig}
\end{figure}

{\it Example 1: Entanglement swapping.} Entanglement swapping (ES)
is a very interesting quantum mechanical phenomenon, which was
originally proposed by \.{Z}ukowski {\em et al.}~\cite{ZZHE},
generalized to multipartite quantum systems by Zeilinger {\em et
al.}~\cite{ZHWZ} and Bose {\em et al.}~\cite{BVK} independently, and
experimentally realized by Pan {\em et al.}~\cite{PBWZ}. The
original ES is based on quantum measurement: Suppose Alice and Bob
share an entangled state, similarly Claire and Danny also share some
entangled states, if Bob and Claire come together and make a
measurement in a suitable basis and communicate their measurement
results classically, then Alice's and Danny's particles may become
entangled. Now we come to use the braiding transformations to
realize the ES. Starting from a separable state
$|0000\rangle_{ABCD}\equiv|0000\rangle_{1234}$, we prepare a state
$|\psi\rangle_{ABCD}$ needed for quantum entanglement swapping due
to the braiding transformations $B_{12}$ and $B_{34}$ as follows:
\begin{eqnarray}\label{state1}
|\psi\rangle_{ABCD}&=&B_{34}B_{12}|0000\rangle_{ABCD}
\nonumber\\
& =&|\Phi^-\rangle_{AB} \otimes |\Phi^-\rangle_{CD}, \\&=&
\frac{1}{\sqrt{2}}(|0 0\rangle-|1 1\rangle)_{AB}\otimes
\frac{1}{\sqrt{2}}(|0 0\rangle-|1 1\rangle)_{CD}, \nonumber
\end{eqnarray}
here for simplicity we have set $\varphi=0$, and $|\Phi^\pm\rangle$
are the usual Bell states. One may verify that
\begin{eqnarray}\label{state2}
|\psi'\rangle_{ABCD}&=&B_{23}B_{34}B_{12}B_{23}|\psi\rangle_{ABCD}\nonumber\\
&=&-|\Phi^-\rangle_{AD} \otimes |\Phi^+\rangle_{BC}, \\&=&
\frac{1}{\sqrt{2}}(-|0 0\rangle+|1 1\rangle)_{AD}\otimes
\frac{1}{\sqrt{2}}(|0 0\rangle+|1 1\rangle)_{BC}, \nonumber
\end{eqnarray}
in other words, after making the successive braiding transformations
$B_{23}B_{34}B_{12}B_{23}$, the entanglement involved in the state
$|\psi\rangle_{ABCD}$ is swapped to $|\psi'\rangle_{ABCD}$,
therefore we have realized the ES (see Fig. 1). The difference
between the original ES scenario and ours is that the former based
on quantum measurement, while the latter based on unitary braiding
transformations without quantum measurement. It is worthy to mention
that the approach of realizing ES by braiding transformations is not
unique. For instance, ES can be done even simpler by using only two
permutations $P_{34} P_{23}$ that acting on the state
$|\psi\rangle_{ABCD}$.

{\it Example 2: Generating the GHZ states and the linear cluster
states.} These are some kinds of important entangled states in
quantum information, such as the well-known GHZ state and the linear
cluster state. (i) It is easy to check that, after acting
$B_{12}B_{23}$ on the initially separable three-qubit state
$|111\rangle_{123}$, one obtains a state
\begin{eqnarray}\label{stateGHZ1}
&&|\psi'\rangle_{GHZ}= B_{12}B_{23}|111\rangle_{123}
\\
&&=\frac{1}{2}(|100\rangle_{123}+|010\rangle_{123}+|001\rangle_{123}+|111\rangle_{123}),\nonumber
\end{eqnarray}
which is equivalent to the standard three-qubit GHZ state
$|\psi\rangle_{GHZ}=1/\sqrt{2}(|000\rangle_{123}+|111\rangle_{123})$
up to a local unitary transformation:
\begin{eqnarray}\label{stateGHZ2}
|\psi'\rangle_{GHZ}=U_a\otimes U_b\otimes U_c |\psi\rangle_{GHZ},
\end{eqnarray}
where $U_a=U_b=U_c =V$, and
\begin{equation}\label{stateGHZ3}
V=\frac{1}{\sqrt{2}}\left(\begin{array}{cc} 1 & 1\\
-1 &1
\end{array}\right)=
\left(\begin{array}{cc} 1 & 0\\
0 & -1
\end{array}\right)\cdot
\frac{1}{\sqrt{2}}\left(\begin{array}{cc} 1 & 1\\
1 & -1
\end{array}\right),
\end{equation}
i.e., the unitary transformation $V$ is decomposed as a product of
the Hadamard gate and the phase gate of $\sigma^z$. In general, one
may obtain the $N$-qubit GHZ states by acting $B_{12}B_{23}\cdots
B_{N-1, N}$ on the initially separable $N$-qubit state
$|11\cdots1\rangle_{12\cdots N}$. (ii) The linear cluster state is
the highly entangled multiparticle state on which one-way quantum
computation is based \cite{Raussendorf}\cite{Prevedel}. The linear
cluster state is locally equivalent to the $N$-qubits ring cluster
state. The random quantum measurement error can be overcome by
applying a feed-forward technique, such that the future measurement
basis depends on earlier measurement results. This technique is
crucial for achieving deterministic quantum computation once a
cluster state is prepared. For four qubits, the linear cluster state
reads
\begin{eqnarray}\label{state3}
|\psi\rangle_{cluster}&=& \frac{1}{2}(
|0\rangle_1|0\rangle_2|0\rangle_3|0\rangle_4 +
|0\rangle_1|0\rangle_2|1\rangle_3|1\rangle_4 +\nonumber\\&&
|1\rangle_1|1\rangle_2|0\rangle_3|0\rangle_4 -
|1\rangle_1|1\rangle_2|1\rangle_3|1\rangle_4).
\end{eqnarray}
However, it is not easy to generate $|\psi\rangle_{cluster}$ by
using only one kind of unitary braiding transformations $B_{i,i+1}$.
In the following, starting from the initial separable four-qubit
state $|0000\rangle_{1234}$, we would like to mathematically
generate the four-qubit linear cluster state by combined using two
kinds of unitary braiding transformations $B_{i,i+1}$ and ${\cal
P}_{i,i+1}$, namely
\begin{eqnarray}\label{state4}
&&|\psi\rangle_{cluster}=P_{23}{\cal
P}_{23}B_{34}B_{12}|0000\rangle_{1234},
\end{eqnarray}
where the phases in ${\cal P}_{23}$ are chosen as $\xi_{00}=0$,
$\xi_{01}=\xi_{10}=\xi_{11}=\pi$, and $P_{23}$ is the usual
permutation operator in Eq. (\ref{permutation}). Moreover, one can
mathematically generate 16 orthogonal four-qubit linear cluster
states by acting $P_{23}{\cal P}_{23}B_{34}B_{12}$ on the initial
states $|ijkl\rangle_{1234}$, where $i,j,k,l$ run from 0 to 1.

Significantly such realizations of entanglement swapping as well as
the GHZ states are purely based on one kind of braiding
transformations $B_{i,i+1}$. Eqs. (\ref{state1})-(\ref{stateGHZ3})
are hopeful to provide an alternative approach for the experimenter
to realize the ES and also generate the GHZ states through a network
of quantum logic gates in the future. Recent realization of the
linear cluster states is based on quantum measurements
\cite{Prevedel}. By using two kinds of braiding transformations, Eq.
(\ref{state4}) has mathematically produced the state
$|\psi\rangle_{cluster}$. Since $B_{i,i+1}$ and ${\cal P}_{i,i+1}$
do not have the same eigenvalues and they cannot be the matrices
representing exchanges within the same braid group representation,
there is still a distance between the mathematical realization Eq.
(\ref{state4}) and the actual physical realization.

\section{$R$-matrix, Hamiltonian and Berry phase in entanglement space}

In Ref. \cite{Kauffman1}, the unitary matrix
$\check{R}_{i,i+1}(\theta,\varphi)$ has been introduced from the
Yang--Baxterization approach \cite{Jimbo} in order to include the
general discussion of the nonmaximally entangled states. To make the
paper be self-contained, we briefly review it in the following.

The Yang-Baxterization of the unitary braiding operator $B_{i,i+1}$
is
 \eqa
\label{belltype1} &&\check{R}_{i,i+1}(x) =\frac{1}{\sqrt{1+x^2}}
 (B_{i,i+1} + x B_{i,i+1}^{-1}),
 \ena
namely, $\check{R}_{i,i+1}(x)$-matrix is a linear superposition of
matrices $B_{i,i+1}$ and $B_{i,i+1}^{-1}$, where $B^{-1}=B^{\dag}$
is the inverse matrix of $B$. The unitary $\check{R}$-matrix is a
generalization of the unitary braiding matrix $B_{i,i+1}$, which
satisfies the Yang--Baxter equation: \eq \label{qybe}
 \check{R}_i(x)\,\check{R}_{i+1}(xy)\,\check{R}_i(y)=
 \check{R}_{i+1}(y)\,\check{R}_i(xy)\,\check{R}_{i+1}(x),\en
where $x$ and $y$ are called the spectral parameters. The braid
relations (\ref{braid}) can be viewed as an asymptotic behavior of
the Yang--Baxter equation. By introducing the new variables of
angles $\theta$ as $
 \cos\theta=(1-x)/{\sqrt{2(1+x^2)}}$, $ \sin\theta= (1+x)/ {\sqrt{2(1+x^2)}}$,
the matrix $\check{R}_{i,i+1}(x)$ may be recast to
$\check{R}_{i,i+1}(\theta,\varphi) = \sin\theta \;{\bf 1}_i \otimes
{\bf 1}_{i+1}+ \cos\theta \;M_{i,i+1}.$
where
$M_{i,i+1}=e^{-i\varphi}
S^+_i \otimes S^+_{i+1}- e^{i\varphi} S^-_i \otimes S^-_{i+1} +
S^+_i \otimes S^-_{i+1}-S^-_i \otimes S^+_{i+1}$,
and $S^\pm=S^x\pm i S^y$ are the matrices for spin-1/2 angular
momentum operators.

 Similar to Eq. (\ref{bmB}), when the unitary
matrix $\check{R}_{i,i+1}(\theta,\varphi)$ acts on the
direct-product states $|kl\rangle$, it is expected to produce the
nonmaximally entangled states as
 \eq \label{Rtrans} \check{R}_{i,i+1}(\theta,\varphi) \left(\begin{array}{l}
  |0 0\rangle
\\  |0 1\rangle
 \\  |1 0\rangle
\\  |1 1\rangle \end{array}\right)=
 \left(\begin{array}{l}
 \sin\theta |0 0\rangle- e^{i\varphi}\cos\theta |1 1\rangle
\\  \sin\theta |0 1\rangle - \cos\theta |1 0\rangle
 \\  \cos\theta |0 1\rangle +\sin\theta|1 0\rangle  )
\\ e^{-i\varphi} \cos\theta|0 0\rangle +\sin\theta |1 1\rangle
  \end{array}\right). \en
Remarkably, the four states in the right-hand side of Eq.
(\ref{Rtrans}) possess the same degree of entanglement (or the
concurrence \cite{98Wo}) equals to $|\sin(2\theta)|$. When
$\theta=\pi/4$, they reduce to the four Bell basis and
correspondingly the matrix $\check{R}_{i,i+1}(\theta,\varphi)$
reduces to the braiding operator $B_{i,i+1}$.

There are two parameters $\theta$, $\varphi$ in the unitary matrix
$\check{R}_{i,i+1}(\theta,\varphi)$. If let $\theta$ be
time-dependent while $\varphi$ be time-independent, one can
construct a Hamiltonian as in Ref. \cite{Kauffman1}. However, the
eigenstates of such a Hamiltonian are separable states, which do not
allow us to study the Berry phases for entangled states. To reach
this purpose, in this paper we will let $\varphi=\omega t$ be
time-dependent while $\theta$ be time-independent.

Equation (\ref{Rtrans}) can be abbreviated as
$\check{R}_{i,i+1}(\theta,\varphi) |\psi(\pi/2,
0)\rangle=|\psi(\theta,\varphi)\rangle$. Taking the Schr{\" o}dinger
equation $i\hbar\partial |\psi(\theta,\varphi)\rangle/\partial t =
H(\theta,\varphi) |\psi(\theta,\varphi)\rangle$ into account, one
obtains $i \hbar
\partial/\partial t [\check{R}_{i,i+1}(\theta,\varphi) |\psi(\pi/2, 0)\rangle]=i
\hbar
\partial/\partial t[|\psi(\theta,\varphi)\rangle] =
H(\theta,\varphi) |\psi(\theta,\varphi)\rangle=H(\theta,\varphi)
\check{R}_{i,i+1}(\theta,\varphi)|\psi(\pi/2, 0)\rangle$.
Now let the parameters $\theta$ be time-independent and
$\varphi(t)=\omega t$, one may arrive at a Hamiltonian through the
unitary transformation $\check{R}_{i,i+1}(\theta,\varphi)$ as
\begin{eqnarray}\label{hamiltonian}
&&H(\theta,\varphi)=i\hbar\frac{\partial
\check{R}_{i,i+1}(\theta,\varphi)}{\partial
t}\check{R}^\dag_{i,i+1}(\theta,\varphi).
\end{eqnarray}
More precisely, the Hamiltonian reads
 \eqa \label{hamiltonian1}
H(\theta,\varphi)&=& \hbar {\dot \varphi} \cos\theta
\left(\begin{array}{cccc}
\cos\theta & 0 & 0 & e^{-i\varphi}\sin\theta \\
0 & 0&  0 & 0 \\
0 & 0 & 0 & 0 \\
e^{i\varphi}\sin\theta& 0 & 0 & -\cos\theta \end{array} \right),
\ena or, $H(\theta,\varphi)= \hbar {\dot \varphi} \cos\theta [
\cos\theta (S^z_i \otimes {\bf 1}_{i+1}+ {\bf 1}_{i} \otimes
S^z_{i+1}) +
  \sin\theta (e^{-i\varphi} S^+_i \otimes S^+_{i+1}+ e^{i\varphi} S^-_i
\otimes S^-_{i+1}]$. In the standard basis
$\{|00\rangle,|01\rangle,|10\rangle,|11\rangle\}$, one observes that
$H(\theta,\varphi)$ has contributions merely on $\{|00\rangle,
|11\rangle\}$, i.e., it makes four-dimensions ``collapse" to
two-dimensions since $\theta$ is assumed to be time-independent.
In the basis of $\{|01\rangle, |10\rangle\}$, the two eigenstates
$|\chi_{01}\rangle=|01\rangle$, $|\chi_{10}\rangle=|10\rangle$ are
degenerate with zero eigenvalues $E_{01}=E_{10}=0$, they will not
give rise to Berry phases so we would not like to discuss them here.
In the basis of $\{|00\rangle, |11\rangle\}$, the two eigenvalues
$E_{\pm}=\pm \hbar {\dot \varphi} \cos\theta$ with two corresponding
eigenstates read
\begin{eqnarray}\label{chistate}
&& |\chi_{+}(\theta,\varphi)\rangle=\cos\frac{\theta}{2} |0
0\rangle+ e^{i\varphi}\sin\frac{\theta}{2} |1
1\rangle,\nonumber\\
&& |\chi_{-}(\theta,\varphi)\rangle=-
e^{-i\varphi}\sin\frac{\theta}{2} |0 0\rangle
+\cos\frac{\theta}{2}|1 1\rangle.\end{eqnarray} Interestingly, the
interval between $E_+$ and $E_-$ depends on $\theta$ that related to
the degree of entanglement of the states. According to Berry's
theory \cite{Wilczek}, when $\varphi(t)$ evolves adiabatically from
0 to $2\pi$, the corresponding Berry phases for the entangled states
are
\begin{eqnarray}\label{holo}
 \gamma_\pm & =& i \int_0^T dt \left\langle
\chi_{\pm}(\theta,\varphi)\right|\frac{\partial}{\partial t}
\left|\chi_{\pm}(\theta,\varphi)\right\rangle= \mp \frac{\Omega}{2},
\end{eqnarray}
 where $\Omega=2\pi(1-\cos\theta)$ is
the familiar solid angle enclosed by the loop on the Bloch sphere
(see Fig. 2).

\begin{figure}
  \includegraphics[width=4cm]{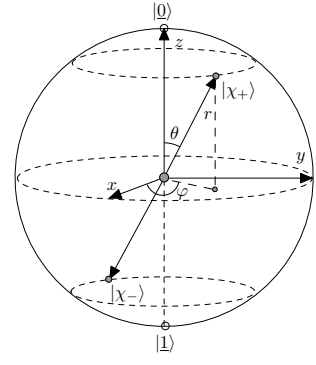}\\
\caption{Berry phases in Bloch space (or the entanglement space).
The parameter $\theta$ comes from the Yang--Baxterization of the
unitary braiding operators, while parameters $\varphi$ originates
from the $q$-deformation of the braiding operators. They define a
point on the unit three-dimensional sphere named the Bloch sphere,
and have definite geometric meanings as angles of longitude and
latitude respectively. Let $\theta$ be time-independent, when the
parameter $\varphi(t)$ evolves adiabatically from 0 to $2\pi$, one
obtains the Berry phases for $\chi_{\pm}(\theta,\varphi)$ as shown
in Eq. (\ref{holo}). The relation between Berry phases and
concurrence of the entangled states $\chi_{\pm}(\theta,\varphi)$ is
$\gamma_\pm=\mp \pi(1-\sqrt{1-{\cal C}^2})$, where ${\cal
C}=|\sin\theta|$ is the concurrence.
}\label{blochfig}
\end{figure}

Actually, the eigenstates $|\chi_{\pm}(\theta,\varphi)\rangle$ are
the $SU(2)$ spin coherent states. If we express the Hamiltonian in
terms of $SU(2)$ generators as \cite{87CSS}
\eqa \label{hamiltonian2} H(\theta,\varphi)= X_1 J_1 + X_2 J_2 +X_3
J_3,\ena
where $X_1=2\hbar {\dot \varphi} \cos\theta \sin\theta
\cos\varphi$, $X_2=2\hbar {\dot \varphi} \cos\theta \sin\theta
\sin\varphi$, $X_3=2\hbar {\dot \varphi} \cos\theta \cos\theta$, and
the $SU(2)$ generators are
\begin{eqnarray}\label{su2}
&&J_1=  (S^+_i \otimes S^+_{i+1}+ S^-_i \otimes S^-_{i+1})/2,
\nonumber\\
&&J_2=  (S^+_i \otimes S^+_{i+1}- S^-_i \otimes S^-_{i+1})/{2i},
\nonumber\\
&&J_3=(S^z_i \otimes {\bf 1}_{i+1}+ {\bf 1}_{i} \otimes
S^z_{i+1})/2,
\end{eqnarray}
based on which one can verify directly that
\begin{eqnarray}\label{su2cohe1}
&&|\chi_{+}(\theta,\varphi)\rangle=\exp[\zeta J_+- \zeta^* J_-]\;
|00\rangle,\nonumber\\&& |\chi_{-}(\theta,\varphi)\rangle=\exp[\zeta
J_+- \zeta^* J_-]\; |11\rangle,
\end{eqnarray}
where $\exp[\zeta J_+- \zeta^* J_-]$ is the spin coherence operators
(and also the usual $D^{\frac{1}{2}}(\theta,\varphi)$-matrix in the
angular momentum theory),
 $J_\pm=J_1 \pm i J_2$ and $\zeta=e^{-i\varphi}
\theta/2$. Berry phase for spin coherence states has been discussed
in \cite{87CSS}, where the corresponding result coincides with Eq.
(\ref{holo}).

\section{Conclusion and Discussion}
In summary, we have shown that braiding transformation is a natural
approach to describe quantum entanglement, by applying the unitary
braiding operators to realize entanglement swapping and to generate
the GHZ states as well as the linear cluster states. The unitary
braiding matrix $B_{i,i+1}$ describes the Bell states and the
Yang--Baxter matrix $\check{R}_{i,i+1}(\theta,\varphi)$ describes
generally entangled states with arbitrary degree of entanglement.
Varying the parameter $\theta$ continuously from 0 to $2\pi$, one
may obtain an ``oscillating entanglement" phenomenon for the
entangled states. A Hamiltonian is constructed from the unitary
$\check{R}_{i,i+1}(\theta,\varphi)$-matrix, where $\varphi=\omega t$
is time-dependent while $\theta$ is time-independent. This in turn
allows us to investigate the Berry phases for the entangled states
in the entanglement space.


Let us make two discussions to end this paper.

(i) Very recently, geometric phases for mixed states \cite{Sjoqvist}
have been observed in experiments by using NMR interferometry
\cite{Du} as well as single photon interferometry \cite{Ericsson}.
Under a certain noisy environment, the states
$|\chi_{\pm}(\theta,\varphi)\rangle$ may become mixed states as
\begin{eqnarray}\label{eqn2}
 \rho_{\pm}(r, \theta,\varphi)&=&r \;|\chi_{\pm}\rangle\langle
\chi_{\pm}|+ (1-r)\rho_{\rm noise},
\end{eqnarray}
where $0 \le r \le 1$. Usually, $\rho_{\rm noise}$ is chosen as
${\bf 1}_i \otimes {\bf 1}_{i+1}/4=(|00\rangle\langle
00|+|01\rangle\langle 01|+|10\rangle\langle 10|+|11\rangle\langle
11|)/4$ and $\rho_{\pm}(r, \theta,\varphi)$ become the generalized
Werner states \cite{Niels}. Following Ref. \cite{Singh}, one may
calculate the geometric phases for the mixed states $\rho_{\pm}(r,
\theta,\varphi)$, however, the computation becomes complicated since
$\rho_{\pm}(r, \theta,\varphi)$ have two nonzero degenerate
eigenvalues in the subspace spanned by $\{ |01\rangle, |10\rangle
\}$. Geometric phases for degenerate mixed states are complicated
and we will discuss them elsewhere. In the following, we would like
to discuss a more simpler case for geometric phases of mixed states,
by restricting the noise in the subspace spanned by $\{ |00\rangle,
|11\rangle \}$. The analysis on such a restriction to the noisy
environment is limited, for it assumes that the states $|01\rangle$
and $|10\rangle$ are decoupled, and the environment only affects the
$|00\rangle$ and $|11\rangle$ subspace.

For simplicity, let us denote $|\underline 0 \rangle\equiv |0
0\rangle$, $|\underline 1 \rangle\equiv |1 1\rangle$, then the
Hamiltonian can be rewritten in a very simple form as
$H(\theta,\varphi)=\hbar {\dot \varphi} \cos\theta\; \hat{\bf r}
\cdot {\mathbf \sigma}$,
where $\hat{\bf r}=(\sin\theta \cos\varphi,\sin\theta
\sin\varphi,\cos\theta)$ is a unit vector on the Bloch sphere, and
${\mathbf \sigma}=(\sigma_1, \sigma_2, \sigma_3)$ is the Pauli
matrix vector in the basis of $\{|\underline 0 \rangle, |\underline
1 \rangle\}$, namely, $\sigma_1=|\underline 0 \rangle\langle
\underline 1| +|\underline 1 \rangle\langle \underline 0|$,
$\sigma_2=-i |\underline 0 \rangle\langle \underline 1| +
i|\underline 1 \rangle\langle \underline 0|$, $\sigma_3=|\underline
0 \rangle\langle \underline 0|-|\underline 1 \rangle\langle
\underline 1|$.
Based on which, the pure states $|\chi_{\pm}(\theta,\varphi)\rangle$
can be rewritten in a density matrix form as $|\chi_{\pm}\rangle
\langle\chi_{\pm}|=(1\!\! 1 \pm \hat{\bf r} \cdot {\mathbf
\sigma})/2,$
where $1\!\! 1=|\underline 0 \rangle\langle \underline
0|+|\underline 1 \rangle\langle \underline 1|$.
In other words, in the basis of $\{|\underline 0 \rangle,
|\underline 1 \rangle\}$, $|\chi_{\pm}\rangle$ may be viewed as
states of a single ``qubit", which allows us to introduce mixed
states and discuss their geometric phases in a particular noisy
environment as follows. By choosing $\rho_{\rm noise}=1\!\! 1 /2$,
one has from Eq. (\ref{eqn2}) that
\begin{eqnarray}\label{rhonew}
&&\rho_{\pm}(r, \theta,\varphi)=\frac{1}{2}(1\!\! 1 \pm {\bf r}
\cdot {\mathbf \sigma}),
\end{eqnarray}
where ${\bf r}=r \hat{\bf r}$. The state $|\chi_{+}\rangle$
corresponds to a point $\hat{\bf r}$
on the Bloch sphere; $\rho_{\rm noise}$ is located on the center of
the Bloch sphere; the unit vector $\hat{\bf r}$ shrinks to ${\bf r}$
when the particular noise is presented and then
$|\chi_{\pm}\rangle\langle \chi_{\pm}|$ turn to be mixed states
$\rho_{\pm}(r,\theta,\varphi)$. Follow the same calculations in
\cite{Singh}, let $r$ and $\theta$ be time-independent, when
parameter $\varphi(t)$ evolves adiabatically from 0 to $2\pi$, one
obtains the geometric phase for the mixed states (\ref{rhonew}) as
\begin{eqnarray}\label{holo1}
\gamma_\pm^{\rm mixed} & = &\mp \arctan (r \tan \frac{\Omega}{2}),
\label{r3}
\end{eqnarray}
which reduces to Eq. (\ref{holo}) for $r=1$.

(ii) The Berry phases in Eq. (\ref{holo}) can be expressed in terms
of the concurrence of the states
$|\chi_{\pm}(\theta,\varphi)\rangle$ as $\gamma_\pm=\mp
\pi(1-\sqrt{1-{\cal C}^2})$, with ${\cal C}=|\sin\theta|$ being the
concurrence. It is well-known that ${\cal C}$ is an invariant of
entanglement for the entangled states
$|\chi_{\pm}(\theta,\varphi)\rangle$, while Berry phase is related
to some certain topological structures. This might bridge a
connection between quantum entanglement and topological quantum
computation. Eventually, when one restricts the discussion to the
basis of $\{|\underline 0 \rangle, |\underline 1 \rangle\}$, by
taking $\theta=\pi/4$, $\phi=-\pi/2$ (or $q=i$), the matrix
$\check{R}_{i,i+1}$ may reduce to the two-dimensional representation
of braiding operators as in Eq. (140) of \cite{Slingerland}, which
has physical applications in non-Abelian quantum Hall systems and
topological quantum field theory.

 {\bf ACKNOWLEDGMENTS}  The authors thank Prof. L. D. Faddeev and
Prof. K. Fijikawa for their encouragement and useful discussions.
This work was supported in part by NSF of China (Grant No. 10575053
and No. 10605013) and Program for New Century Excellent Talents in
University.

\end{document}